\documentclass{PoS}
\usepackage{slashed,wrapfig}
\newcommand{\beqn}{\begin{eqnarray}}
\newcommand{\eeqn}{\end{eqnarray}}
\newcommand{\beqs}{\begin{subequations}}
\newcommand{\eeqs}{\end{subequations}}
\newcommand{\eq}[1]{(\ref{#1})}

\newcommand{\cL}{{\cal L}}

\newcommand{\cM}{{\cal M}}

\title{Can nothing be a superconductor and a superfluid?\footnote{Contribution to the proceedings of the workshop "The many faces of QCD", 2-5 November 2010, Gent, Belgium.}
}

\ShortTitle{Can nothing be a superconductor and a superfluid?}

\author{M. N. Chernodub\thanks{This work was supported by Grant No. ANR-10-JCJC-0408 HYPERMAG.}\thanks{On leave from Institute for Theoretical and Experimental Physics, Moscow, Russia.}\\
CNRS, Laboratoire de Math\'ematiques et Physique Th\'eorique,
Universit\'e Fran\c{c}ois-Rabelais, \\ F\'ed\'eration Denis Poisson - CNRS,
Parc de Grandmont, Universit\'e de Tours, 37200 France\\
Department of Physics and Astronomy, University of Gent, Krijgslaan 281, S9, B-9000 Gent, Belgium\\
E-mail: \email{maxim.chernodub@lmpt.univ-tours.fr}}

\abstract{A superconductor is a material that conducts electric current with no resistance. Superconductivity and magnetism are known to be antagonistic phenomena: superconductors expel weak external magnetic field (the Meissner effect) while a sufficiently strong magnetic field, in general, destroys superconductivity. In a seemingly contradictory statement, we show that a very strong magnetic field can turn an empty space into a superconductor. The external magnetic field required for this effect should be about $10^{16}$  Tesla ($eB \sim 1\,\mbox{GeV}^2$). The physical mechanism of the exotic vacuum superconductivity is as follows: in strong magnetic field the dynamics of virtual quarks and antiquarks is effectively one-dimensional because these electrically charged particles tend to move along the lines of the magnetic field. In one spatial dimension a gluon-mediated attraction between a quark and an antiquark of different flavors inevitably leads to formation of a colorless spin-triplet bound state (a vector analogue of the Cooper pair) with quantum numbers of an electrically charged $\rho^\pm$ meson. Such quark-antiquark pairs condense to form an anisotropic inhomogeneous superconducting state similar to the Abrikosov vortex lattice in a type-II superconductor.  The onset of the superconductivity of the charged $\rho^\pm$ mesons should also induce an inhomogeneous superfluidity of the neutral $\rho^0$ mesons. The vacuum superconductivity should survive at very high temperatures of typical Quantum Chromodynamics (QCD) scale of $10^{12} K$ ($100\ {\mathrm{MeV}}$). We propose the phase diagram of QCD in the plane "magnetic field - temperature".
}

\FullConference{The many faces of QCD\\
		November 2-5, 2010\\
		Gent Belgium}

\begin{document}

\section{The physical mechanism of electric superconductivity of vacuum}

Recently, we suggested that the vacuum in a sufficiently strong magnetic field may undergo a spontaneous transition to an electromagnetically\footnote{We stress the word ``\emph{electromagnetic}'' in order to distinguish the proposed superconducting state of the vacuum from the ``color superconductivity'' (which may exist in a sufficiently dense quark matter) and from the ``dual  superconductivity'' associated with confining features of the gluonic fields in the pure Yang-Mills vacuum.} superconducting state~\cite{ref:I,ref:II}. The effect emerges as a result of an interplay between strong fundamental forces and electromagnetic interactions. Below we discuss a standard mechanism of the conventional superconductivity, and then turn our attention to the suggested superconductivity of empty space.

\subsection{Conventional superconductivity}
\begin{wrapfigure}[15]{r}{85mm}
\begin{center}
\vskip -5mm
\includegraphics[scale=0.7,clip=false]{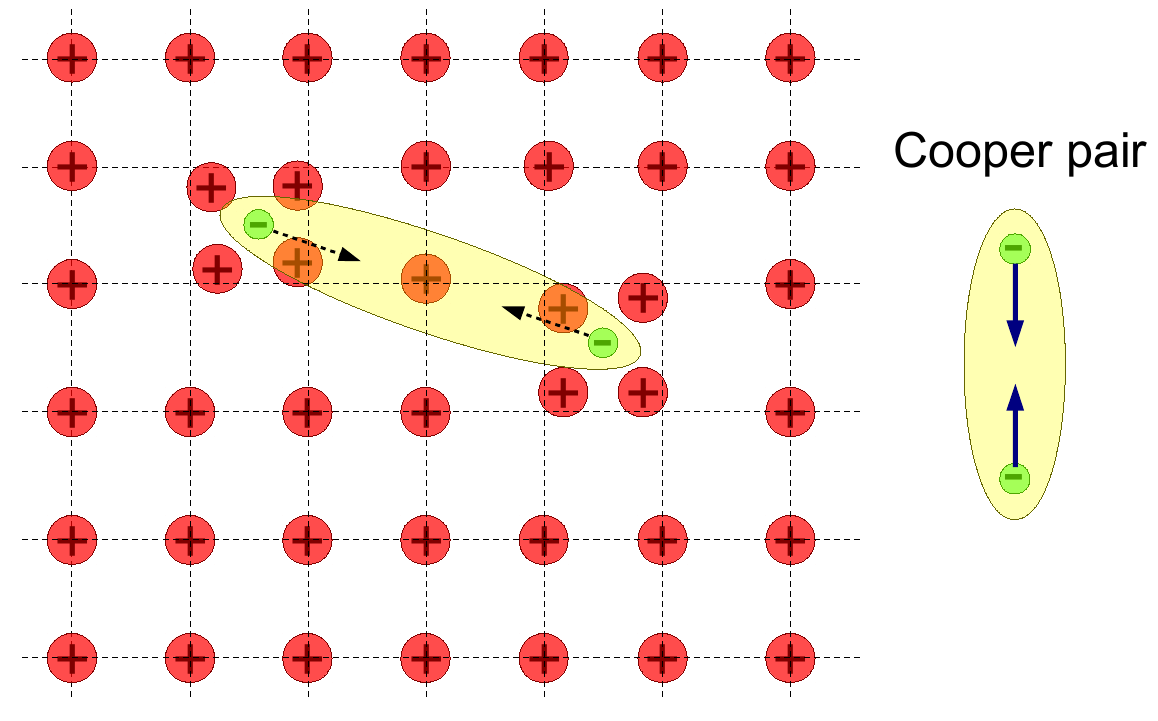}
\end{center}
\vskip -5mm
\caption{Very schematically: the distortion of the ion lattice around one electron attracts another electron, leading to formation of 
the Cooper pair.}
\label{fig:usual}
\end{wrapfigure}
In a conventional superconductor the superconductivity appears as a result of condensation of Cooper pairs. The Cooper pair is a  bound state of two electrons which is formed due to small attractive force between the electrons. The attraction is mediated by a phonon exchange: one electron deforms a lattice of the positively charged ions of a metal, creating a local excess of the positive charge. This excess attracts another electron, and the whole process of the attractive interaction can be viewed as an exchange of collective excitations of the ion lattice (phonons) between the electrons, Figure~\ref{fig:usual}. The formation of the Cooper pairs is promoted by the fact that the dynamics of the electrons near the Fermi surface is basically one-dimensional, and in one spatial dimension the bound states are formed for arbitrarily weak attraction (the Cooper theorem).

\vskip 5mm

There are three important ingredients of the standard mechanism of the Cooper pair formation: 
\begin{itemize}
\item[A)] the presence of carriers of electric charge (of electric current);
\item[B)] the reduction of physics from (3+1) to (1+1) dimensions;
\item[C)] the attractive interaction between the like-charged particles.
\end{itemize}

Below we describe the physical mechanism of the magnetic-field-induced superconductivity of the vacuum and we demonstrate that all these ingredients, A, B and C, are present in our mechanism. 

\clearpage

\subsection{Electromagnetic superconductivity of the vacuum}

\begin{wrapfigure}[11]{r}{55mm}
\begin{center}
\vskip -12mm
\includegraphics[scale=0.45,clip=true]{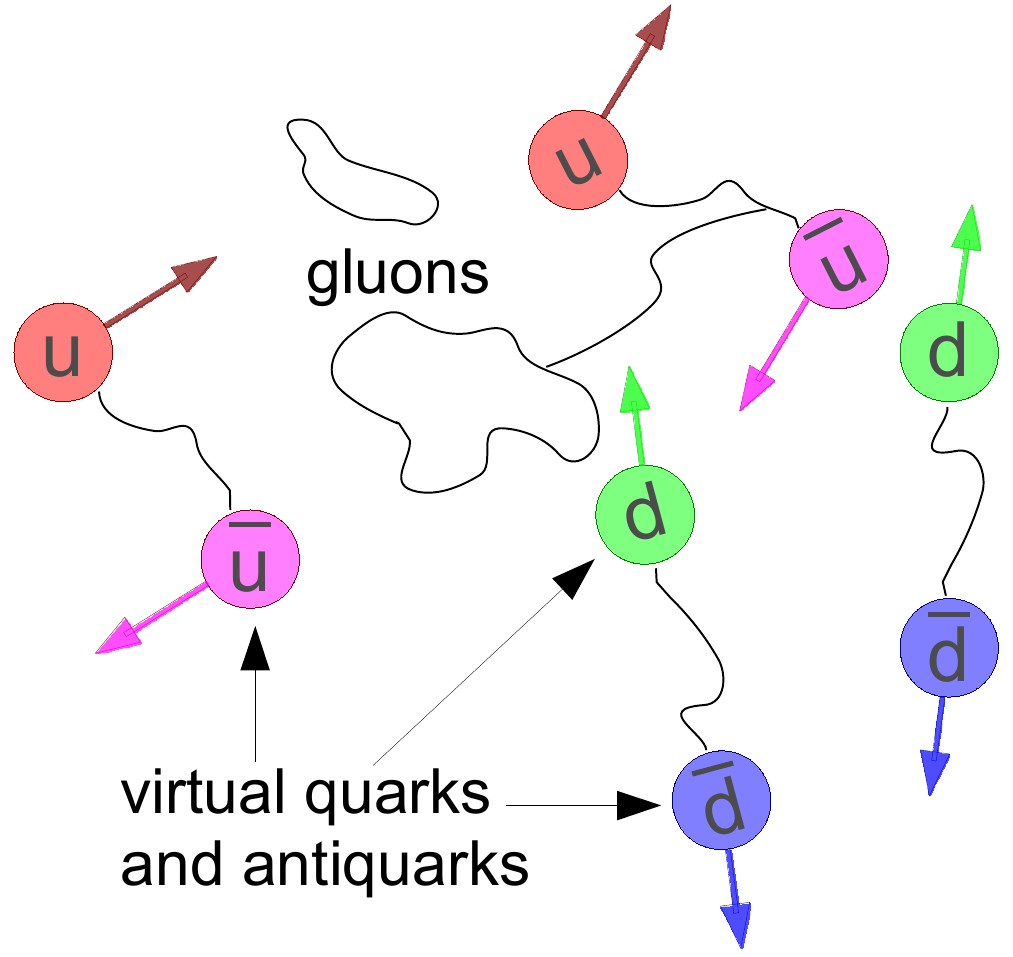} 
\end{center}
\vskip -5mm
\caption{The vacuum of QCD: a boiling soup of virtual quarks and gluons.}
\label{fig:vacuum:nonmagnetic}
\end{wrapfigure}

\noindent {\bf {A. Presence of electric charge carriers.}} \\
In fact, there are no charge carriers in the vacuum in normal conditions. However, the vacuum is a boiling soup of virtual particles (``quantum fluctuations'', Figure~\ref{fig:vacuum:nonmagnetic}), and some of these particles may become real in certain cases. The simplest relevant example is the Schwinger effect: it turns out that it is energetically favorable for the vacuum to produce electron-positron pairs in a uniform time-independent background of a sufficiently strong external electric field~\cite{ref:Schwinger}. Below we show that a similar effect exists in a background of a strong magnetic field which forces the vacuum to develop certain electrically charged condensates of quark-antiquark pairs. The presence of the positively charged condensate automatically implies the presence of a negatively charged condensate of equal magnitude.  As a result, the energy of the vacuum is lowered, while the net electric charge of the vacuum is zero~\cite{ref:I,ref:II}. Despite of the net electric neutrality, the vacuum may (super)conduct since a weak external electric field pushes the positively and negatively charged condensates in opposite directions, thus creating a net electric current along the electric field. 

\vskip 3mm

\noindent {\bf {B. Dimensional reduction.}} \\
It is the presence of the Fermi surface that leads to the effective dimensional reduction of the electron dynamics and facilitates the formation of the Cooper pairs in a conventional superconductor. In the vacuum all chemical potentials are zero and, obviously, Fermi surfaces do not exist. However, a sufficiently strong external magnetic field should reduce the motion of electrical charges: the low-energy particles may move only along the lines of the magnetic field. This effect guarantees the required (3+1) $\to$ (1+1) dimensional reduction.

\vskip 3mm

\noindent {\bf {C. Attractive interaction between the like-charged particles.}}\\
All possible types of virtual particles  "boil" in the quantum  vacuum. The purely electromagnetic sector of the vacuum contains virtual electron-positron pairs and virtual photons, and the strongly interacting sector consists of virtual quark-antiquark pairs and virtual gluons (we ignore heavier particles and weaker interactions). The quarks are electrically charged particles so that their dynamics leaves a trace in the electromagnetic sector as well.

Firstly, let us consider the purely electromagnetic sector described by Quantum Electrodynamics (QED). The electrons repel each other due to the {\it photon} exchange. Obviously, there are no attractive {\it phonon}-like interactions between two electrons in the vacuum and, consequently, the Cooper pairs cannot form (the same statement is valid for positrons). Electron-positron bound states (positronium) are not interesting for us because such states are electrically neutral. This is a simplest reason why the vacuum superconductivity cannot emerge in the pure QED vacuum.

Secondly, we notice that the strongly interacting sector of the vacuum does contain an analogue of the phonon. It is a gluon, a carrier of the strong force. It provides an attractive interaction between the quarks and anti-quarks and binds them in the pairs called mesons. The quarks and antiquarks of the same electric charge can also be bound by the gluon exchange (for example, the $u$ quark with the electric charge $q_u = + 2 e/3$ and the $\bar d$ antiquark with the electric charge $q_{\bar d} \equiv - q_d = + e/3$ are bound by the gluon-mediated interaction, forming the $\rho$ meson with the electric charge $q_\rho = +e$).

\vskip 3mm

\begin{figure}[!thb]
\begin{center}
\includegraphics[scale=0.56,clip=false]{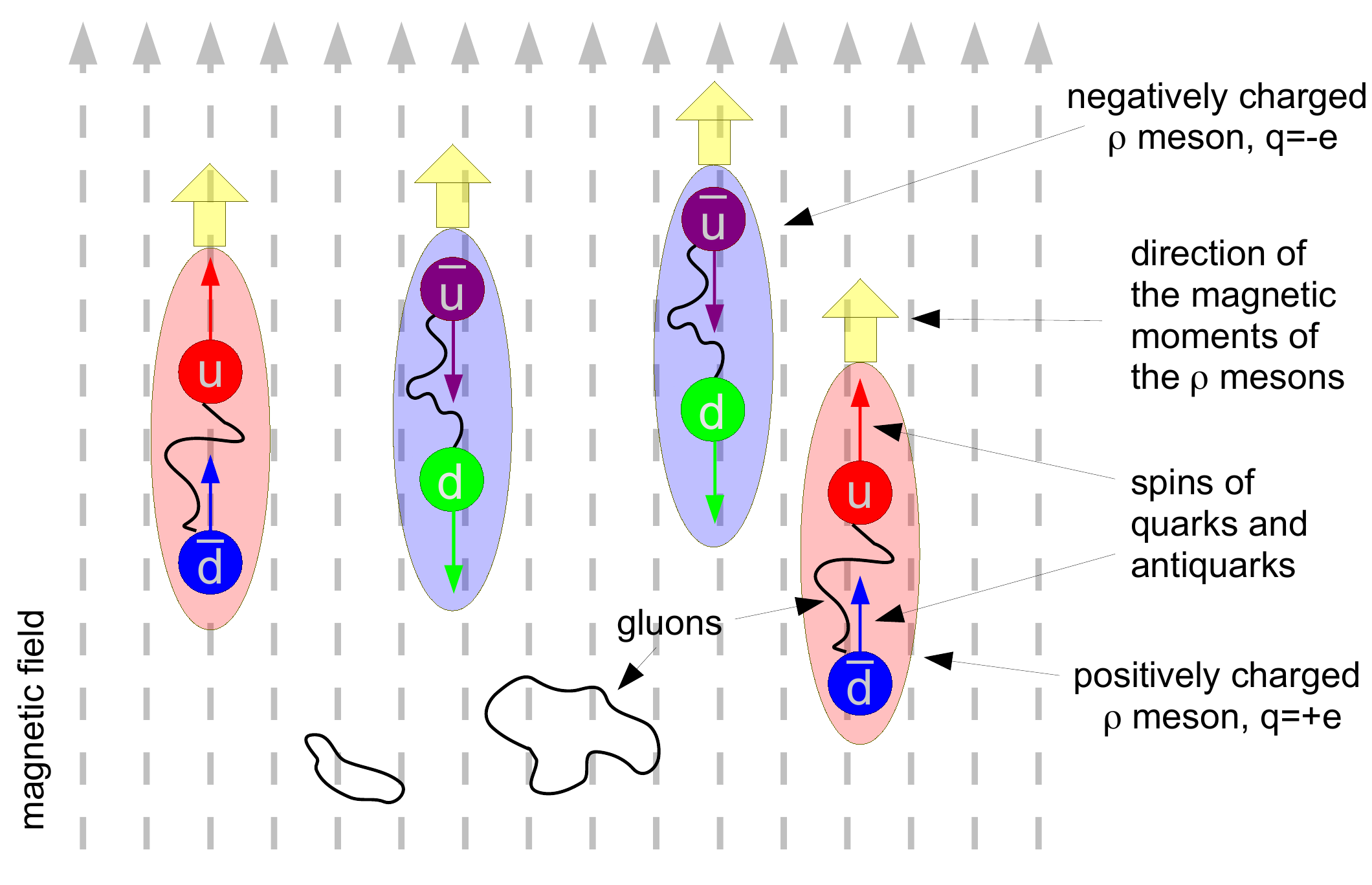}
\end{center}
\vskip -5mm
\caption{Very schematically: formation of quark-antiquark pairs with quantum numbers of $\rho^\pm$ mesons via the gluon exchange  in a background of a very strong magnetic field.  The quarks and, consequently, their bounds states (mesons) may move only along the axis of the magnetic field.}
\label{fig:vacuum:magnetic}
\end{figure}

\noindent {\bf {A+B+C. Stable charged bound states due to gluon interaction and dimensional reduction.}} \\
The physical mechanism of the exotic vacuum superconductivity is as follows: in strong magnetic field the dynamics of virtual quarks and antiquarks is effectively one-dimensional because these electrically charged particles tend to move along the lines of the magnetic field. In one spatial dimension a gluon-mediated attraction between a quark and an antiquark inevitably leads to formation of a quark-antiquark bound state (Figure~\ref{fig:vacuum:magnetic}). This bound state should be: 
\begin{itemize}
\item[1)] a colorless state due to the quark confinement property (otherwise the energy of the bound state would be infinite due to nonperturbative QCD effects);
\item[2)] a state, composed of the quark and the antiquark of different flavors (otherwise the state would be electrically neutral\footnote{If the state is composed of the quark and antiquark of the same flavor then the electric charge of this state is zero.});
\item[3)] a vector (spin-triplet) state (in order to occupy a lowest energy state).
\end{itemize}
The physically interesting bound state has quantum numbers of an electrically charged $\rho^\pm$ meson.

The condensate of the quark-antiquark pairs is an energetically favorable state. Similarly to the Schwinger effect, the charged particles (the quark-antiquark pairs in our case) emerge from the vacuum. In contrast to the Schwinger effect, these pairs form a condensate.  And, in contrast to the virtual nature of the quantum vacuum fluctuations, the condensate of the $\rho$ mesons is a real state.

\subsection{Energetic arguments}

In the absence of the external magnetic field the quark-antiquarks pairs are unstable. For example, the $\rho$ meson has so short lifetime, so that it is often called a ``resonance'', not even a particle. However, even very simplified kinematical arguments of Ref.~\cite{ref:I} suggest that a sufficiently strong magnetic field makes the charged $\rho$ meson stable against known decay modes. Below we give simple energetic arguments in favor of the magnetic-field-induced $\rho$-meson condensation.

Consider a free relativistic spin-$s$ particle moving in a background of the external magnetic field $B$. The energy levels $\varepsilon$ of the particle are:
\beqn
\varepsilon_{n,s_z}^2(p_z) = p_z^2+(2 n - 2 s_z + 1) |eB| + m^2\,,
\label{eq:energy:levels}
\eeqn
where the integer $n\geqslant 0$ labels the energy levels, and other quantities characterize the properties of the pointlike particle: mass $m$, the projection of the spin $s$ on the field's axis $s_z = -s, \dots, s$, the momentum along the field's axis, $p_z$, and the electric charge $e$.

It is clear from Eq.~\eq{eq:energy:levels} that the ground state corresponds to  $p_z=0$, $n_z = 0$ and $s_z = s$. The ``minimal mass'', corresponding to the ground state energy of the charged $\rho$ mesons (with $s=1$) is
\beqn
m_{\rho^\pm}^2(B) = m_{\rho}^2 - e B\,.
\label{eq:m2:pi:B}
\eeqn

Thus, the ground state energy of the charged $\rho$ meson should decrease with the increase of the magnetic field $B$. When the magnetic field reaches the critical strength,
\beqn
B_c = m_\rho^2/e \approx 10^{16}\,\mbox{Tesla}\,,
\label{eq:eBc}
\eeqn
the ground state energy of the $\rho^\pm$ mesons becomes zero ($m_\rho = 775.5\,\mbox{MeV}$ is the $\rho$-meson mass). 

As the field increases above the critical value~\eq{eq:eBc}, the ground state energy of the charged $\rho$ mesons becomes purely imaginary thus signaling a tachyonic instability of the QCD ground state. At $B>B_c$ fields the strongly interacting (QCD) sector of the vacuum spontaneously develops the $\rho$-meson condensates. Since the condensed mesons are electrically charged, their condensation implies, almost automatically, an electromagnetic superconductivity of the condensed state~\cite{ref:I,ref:II}.

The suggested condensation of the charged $\rho$ mesons is similar to the Nielsen-Olesen instability of the gluonic vacuum in Yang-Mills theory~\cite{ref:NO}, and to the magnetic-field-induced Ambj\o rn--Olesen condensation of the $W$-bosons in the standard electroweak model~\cite{ref:AO}. Both the $\rho$ mesons in QCD, the gluons in Yang-Mills theory, and the $W$ bosons in the electroweak model have the anomalously large gyromagnetic ratio, $g=2$ [explicitly implemented in Eq.~\eq{eq:energy:levels}] which supports the mentioned effects. The value $g=2$ for the $\rho$ mesons is supported by various arguments~\cite{ref:Sakurai,ref:g2}.

In contrast to the gluons and $W$ bosons, the $\rho$ meson is a composite particle which, in the absence of the external field, has a finite radius of the order of a typical QCD scale, $r_\rho \simeq 0.5 \, {\mathrm{fm}} \simeq (400\ { \mathrm{MeV}})^{-1}$. The radii of the lowest Landau levels of the $u$ and $d$ quarks at the  critical field~\eq{eq:eBc} are $r_{u,c} \simeq 0.32\,{\mathrm{fm}}$ and $r_{d,c} \simeq0.45\,{\mathrm{fm}}$, respectively, which are comparable with the the radius $r_\rho$ of the $\rho$ meson itself. Thus, generally speaking, the $\rho$ meson cannot be treated as a pointlike particle at the required strong magnetic fields.  Nevertheless,  we work with a pointlike description of the $\rho$ mesons using a bosonic effective model (Section~\ref{sec:VDM}) and then we discuss the $\rho$-meson condensation beyond the pointlike approximation using a more fundamental Nambu--Jona-Lasinio (NJL) model~\cite{ref:NJL} (Section~\ref{sec:NJL}). We show that both approaches give qualitatively the same results.

\clearpage
\section{Electromagnetic superconductivity of vacuum in vector meson dominance model}
\label{sec:VDM}

A simple realization of the magnetic-field-induced superconductivity~\cite{ref:I} can be found in a quantum electrodynamics for the $\rho$ mesons based on the vector meson dominance property~\cite{ref:Sakurai,ref:QED:rho}:
\beqn
{\cal L} {=} -\frac{1}{4} \ F_{\mu\nu}F^{\mu\nu}
{-} \frac{1}{2} (D_{[\mu,} \rho_{\nu]})^\dagger D^{[\mu,} \rho^{\nu]} + m_\rho^2 \ \rho_\mu^\dagger \rho^{\mu}
{-}\frac{1}{4} \ \rho^{(0)}_{\mu\nu} \rho^{(0) \mu\nu}{+}\frac{m_\rho^2}{2} \ \rho_\mu^{(0)}
\rho^{(0) \mu} +\frac{e}{2 g_s} \ F^{\mu\nu} \rho^{(0)}_{\mu\nu}\,, 
\nonumber
\eeqn
where $D_\mu = \partial_\mu + i g_s \rho^{(0)}_\mu - ie A_\mu$ is the covariant derivative,
$g_s \equiv g_{\rho\pi\pi} \approx 5.88$ is the $\rho\pi\pi$ coupling,
$A_\mu$ is the photon field with the field strength $F_{\mu\nu} = \partial_{[\mu,} A_{\nu]}$, the fields
$\rho_\mu = (\rho^{(1)}_\mu - i \rho^{(2)}_\mu)/\sqrt{2}$ and $\rho^{(0)}_\mu \equiv \rho^{(3)}_\mu$
are, respectively, the fields of the charged and neutral vector mesons with the mass~$m_\rho$, 
and $\rho^{(0)}_{\mu\nu} = \partial_{[\mu,} \rho^{(0)}_{\nu]} - i g_s \rho^\dagger_{[\mu,} \rho_{\nu]}$.
The model possesses the electromagnetic $U(1)$ gauge invariance:
\beqn
U(1)_{\mathrm{e.m.}}: 
\qquad
\rho_\mu(x) \to e^{i \omega(x)} \rho_\mu(x)\,,
\qquad
A_\mu(x) \to A_\mu(x) + \partial_\mu \omega(x)\,.
\label{eq:gauge:invariance}
\eeqn
The last term in the above Lagrangian describes a nonminimal coupling of the $\rho$ mesons to the electromagnetic field $A_\mu$, implying the anomalous gyromagnetic ratio ($g = 2$) of the charged $\rho^\pm$ mesons. The last term plays a crucial role in the mechanism of the vacuum superconductivity.

\begin{wrapfigure}[11]{r}{70mm}
\begin{center}
\vskip -3mm
\includegraphics[scale=0.45,clip=false]{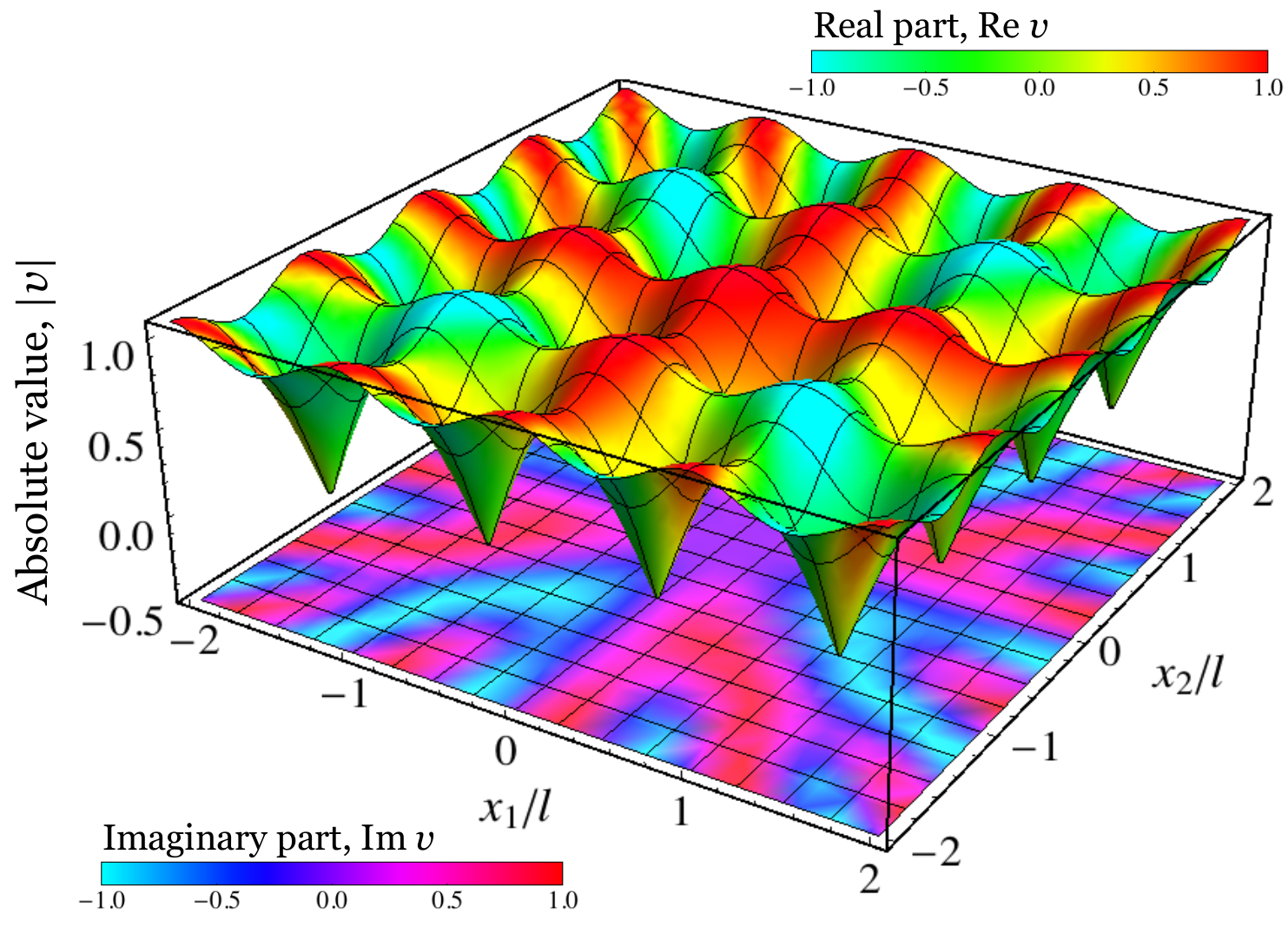}
\end{center}
\vskip -5mm
\caption{Typical structure of the superconducting condensate $v = (\rho_1 + i \rho_2)/2$ in the $(x_1,x_2)$ plane ($l \equiv L_B = \sqrt{2\pi/|eB|}$), perpendicular to the background magnetic field $\vec B = (0,0,B)$.}
\label{fig:condensate}
\end{wrapfigure}
The quadratic part of the energy density,
\beqn
\epsilon_0^{(2)}(\rho_\mu) & = & \sum_{i,j=1}^2\rho_i^\dagger \cM_{ij} \rho_j + m_\rho^2 (\rho_0^\dagger \rho_0 + \rho_3^\dagger \rho_3)\,, \nonumber\\
\cM & = &
\left(
\begin{array}{cc}
m_\rho^2 & i e B \\
- i e B & m_\rho^2
\end{array}
\right)\,,
\label{eq:cM}
\eeqn
shows that the mass terms for $\rho_0$ and $\rho_3$ components are diagonal and their prefactors $m_\rho^2$ are unaltered by the external magnetic field\footnote{Notice that in a dense isospin--asymmetric matter the longitudinal condensates (with $\rho_{0,3} \neq 0$) do emerge~\cite{ref:isospin}, leading to an electromagnetically superconducting state~\cite{Ammon:2008fc}. On the contrary, our $\rho$-meson condensates have a spatially-transverse structure (with $\rho_{1,2} \neq 0$), and they appear in completely empty space.}. However, the Lorentz components $\rho_1$ and $\rho_2$ possess the non--diagonal mass matrix $\cM$. The eigenvalues $\mu_{\pm}$ and the corresponding eigenvectors $\rho_{\pm}$ of the mass matrix ${\cal M}$ are, respectively, as follows:\\
\parbox[l]{8cm}{
\beqn
\mu_{\pm}^2 = m_\rho^2 \pm e B\,,
\quad
\rho_{\pm} = \frac{1}{\sqrt{2}} (\rho_1 \mp i \rho_2)\,.
\nonumber
\eeqn
}\\
The $\rho_-$ state with the mass~\eq{eq:m2:pi:B} becomes unstable against the condensation if the magnetic field exceeds the critical value~\eq{eq:eBc}. The emerging condensate has a typical structure of the Abrikosov lattice~\cite{Abrikosov:1956sx}, composed of the new topological objects in QCD, the ``$\rho$ vortices''~\cite{ref:I}, Figure~\ref{fig:condensate}.

The condensation of the electrically charged particles leads to the electric superconductivity, which, in the case of the charged $\rho^\pm$ mesons is accompanied by superfluidity of the neutral $\rho^{(0)}$ mesons~\cite{ref:I}. Thus, our calculations suggest that the external magnetic field, $B > B_c$, induces both superconductivity and superfluidity of the empty space.

\clearpage
\section{Electromagnetic superconductivity of vacuum in Nambu--Jona-Lasinio model}
\label{sec:NJL}

The signatures of the vacuum superconductivity in strong magnetic field can also be found in an extended two-flavor ($N_f = 2$) Nambu-Jona-Lasinio model with three colors ($N_c = 3$) \cite{ref:NJL,Ebert:1985kz}:
\beqn
\cL(\psi,\bar\psi) = \bar \psi \bigl(i \slashed \partial + {\hat Q} \, {\slashed {\cal A}} - \hat M^0 \bigr) \psi 
& & \!\! + \frac{G_S}{2} \bigl[\bigl(\bar \psi \psi\bigr)^2 + \bigl(\bar \psi i\gamma^5 \vec \tau \psi\bigr)^2 \bigl] \nonumber \\
& & - \frac{G_V}{2}  \sum\nolimits_{i=0}^{3} \left[\bigl(\bar \psi \gamma_\mu \tau^i\psi\bigr)^2 + \bigl(\bar \psi  \gamma_\mu \gamma_5  \tau^i \psi\bigr)^2\right]\,,
\label{eq:L:NJL}
\eeqn
where $\hat M^0 = {\mathrm{diag}} (m^0_u, m^0_d)$ is the bare quark mass matrix,  $\hat Q = {\mathrm{diag}} (q_u,q_d) = {\mathrm{diag}} (+2e/3, - e/3)$ is the charge matrix, $\psi = (u,d)^T$ is the doublet of the light quarks, $\vec \tau = (\tau^1,\tau^2,\tau^3)$ are the Pauli matrices, and  $G_S$ and $G_V$ are, respectively, the scalar and vector couplings of four-quark interactions.

The strong magnetic field induces the condensate of the $\rho$-meson field $\rho_\mu = {\bar u} \gamma_\mu d$~\cite{ref:II}:
\beqn
\langle \bar u \gamma_1 d\rangle = - i \langle \bar u \gamma_2 d\rangle = \rho_0 (B) \, K\Bigl(\frac{x_1+ i x_2}{L_B}\Bigr)\,, 
\qquad
K(z) & = & e^{-\frac{\pi}{2} (|z|^2 + {\bar z}^2)} \sum\limits_{n = -\infty}^{+ \infty} c_n e^{- \pi n^2 + 2 \pi n {\bar z}}\,,
\label{eq:ud:cond}
\eeqn
where $c_n$ are complex parameters, $L_B = \sqrt{2 \pi/|e B|}$ is the magnetic length and the complex field $\rho_0 = \rho_0(x^1,x^2)$ determines the magnitude of the  $\rho$-meson condensate:
$\rho_0 = 0$ if $B < B_c$ and
\beqn
\rho_0 (B) = e^{i \theta_0} C_\phi \frac{m_q(B)}{G_V} {\left(1 - \frac{B_c}{B}\right)}^{1/2}\,, \qquad B \geqslant B_c\,.
\label{eq:phi0}
\eeqn
Here $\theta_0$ is an arbitrary coordinate-independent phase, $C_\phi \approx 0.51$ is a constant, and $m_q(B)$ is the quark mass which is a smooth function of the magnetic field $B$. 

The condensate~\eq{eq:ud:cond} has a typical structure of the Abrikosov lattice in a type-II superconductor~\cite{Abrikosov:1956sx}. The ground state of the superconducting vacuum has an inhomogeneous periodic structure made of new type of topological defects, $\rho$ vortices. These vortices are composed of the quark-antiquark vector condensates. The parameters $c_n$ should be fine-tuned 
by the minimization of the energy with the ansatz~\eq{eq:ud:cond}. However, typical lattice conformations in type-II superconductors cost a few percents of total condensation energy, so that the choice $c_n=1$ (which corresponds to the square lattice, Figure~\ref{fig:condensate}) is a good approximation to the true energy minimum.

In order to probe the conducting properties of the condensate~\eq{eq:ud:cond} we apply a weak external electric field $E = (E_x, E_y, E_z)$ to the vacuum, $E \ll B$. One can easily find that at $B < B_c$ the induced electric current $J^\mu(x) = \sum\nolimits_{f=u,d} q_f  \langle \bar \psi_f \gamma^\mu \psi_f\rangle$ is zero, naturally indicating that the vacuum stays in the insulating phase at weak magnetic fields. However, at $B \geqslant B_c$ one gets 

\beqn
\frac{\partial {\mathcal Q}(t,z)}{\partial z} + \frac{\partial {\mathcal J}_z (t,z)}{\partial t} 
= \frac{2 C_q}{(2 \pi)^3} e^3 \bigl(B - B_c\bigr) \, E_z \,, 
\qquad \quad 
{\mathcal J}_x = {\mathcal J}_y = 0\,,
\label{eq:London:local}
\eeqn
where ${\mathcal Q}$ and ${\vec {\mathcal J}} \equiv ({\mathcal J}_x, {\mathcal J}_y, {\mathcal J}_z) $ are, respectively, the electric charge density $Q \equiv J^0$ and electric current density $\vec J = (J_x, J_y, J_z)$, averaged over the perpendicular ($x^1,x^2$) plane, and $C_q \approx 1$. Apart from prefactors, the transport laws in the NJL model~\eq{eq:London:local} and in the $\rho$-meson electrodynamics (Section~\ref{sec:VDM}) are identical. The transport law~\eq{eq:London:local} can be rewritten in a Lorentz-covariant form~\cite{ref:II}.

Equation~\eq{eq:London:local} is a London equation for  an anisotropic superconductivity. Thus, we have just found that the strong magnetic field induces the new electromagnetically superconducting phase of the vacuum if $B>B_c$. An empty space becomes an anisotropic inhomogeneous superconductor.

\clearpage

\section{Phase diagram of finite-temperature QCD in strong magnetic field}

\begin{figure}[!thb]
\begin{center}
\includegraphics[scale=0.56,clip=false]{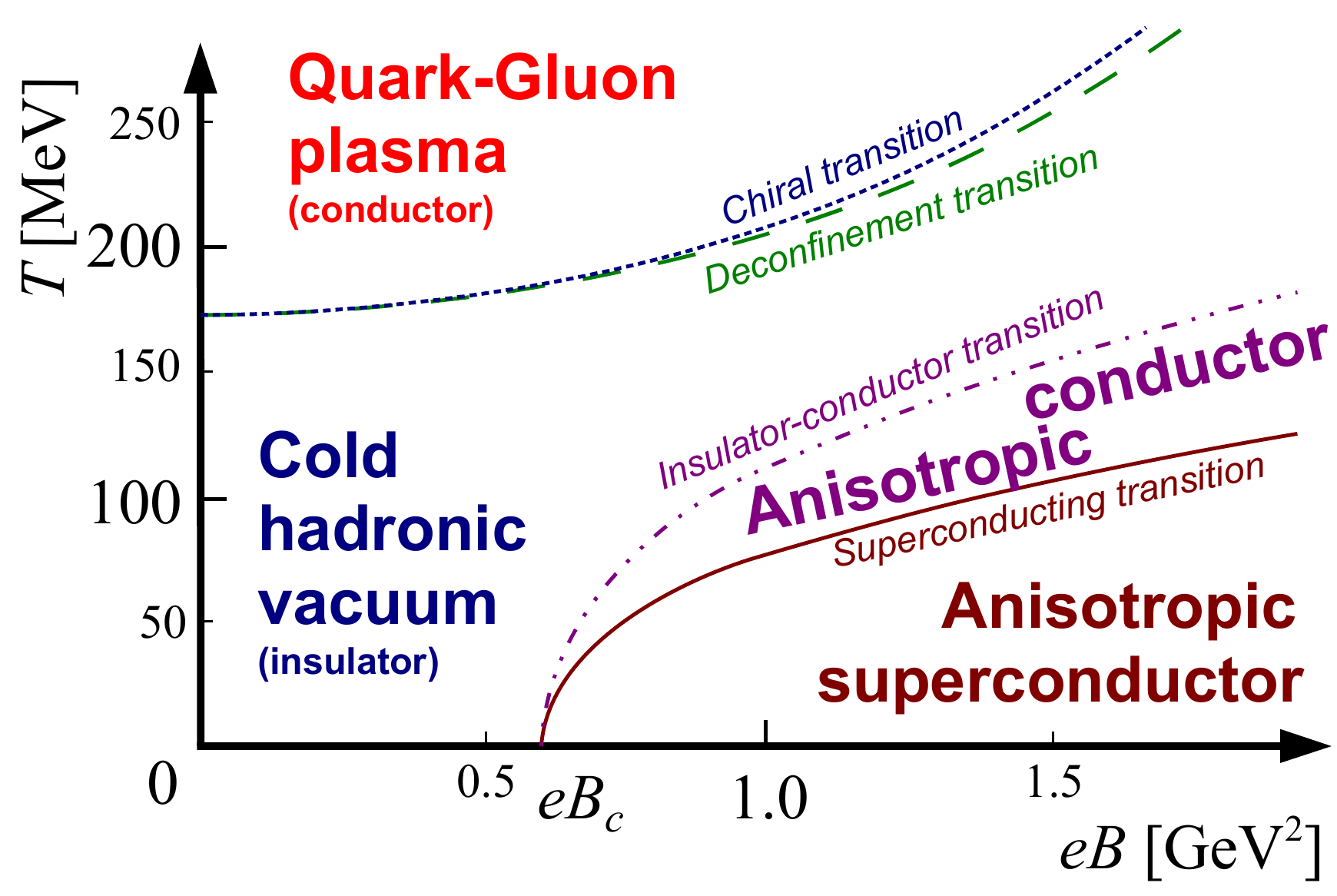}
\end{center}
\vskip -5mm
\caption{Qualitative phase diagram of finite-temperature QCD in strong magnetic field.}
\label{fig:phase}
\end{figure}

We think that the phase diagram of finite-temperature QCD in strong magnetic field should have the structure plotted in Figure~\ref{fig:phase}. The phase diagram has the following specific features:

\begin{enumerate}

\item {\bf Chiral and deconfinement phase transitions.} According to analytical estimates and numerical simulations of lattice QCD, the critical temperatures of deconfinement and chiral transitions are increasing functions of the strength of the magnetic field~\cite{ref:transition}. The behavior of the chiral transition agrees well with the enhancement of the chiral symmetry breaking by the strong magnetic field, known as the magnetic catalysis~\cite{ref:catalysis}. There is also evidence that these transitions may split at finite magnetic field, although the splitting may be very small.

\item {\bf Anisotropically conducting phase.}
At strong values of the magnetic field and, simultaneously, at high temperatures a new, asymmetrically conducting phase may appear. In this phase the vacuum behaves as a conductor along the axis of the magnetic field, and as an insulator in the directions perpendicular to the axis. The presence of this phase was found in lattice simulations of Ref.~\cite{ref:conductivity}.

\item  {\bf Asymmetric superconductor and superfluid.} At low temperatures and high magnetic fields the vacuum behaves as a superconductor and a superfluid, as we discussed above. The vacuum behaves as a superconductor along the axis of the magnetic field, and as an insulator in the perpendicular directions. This phase was suggested in Ref.~\cite{ref:I}, supported by arguments of Ref.~\cite{ref:II} and found numerically in recent simulations of quenched QCD~\cite{ref:lattice}.

\end{enumerate}

Finally, we would like to mention that QCD in Euclidean space at finite magnetic field has no sign problem of the fermionic determinant contrary to the finite-density QCD. Thus, verification of the predicted phase diagram, Figure~\ref{fig:phase}, should not encounter principal obstacles in numerical lattice simulations.

\clearpage

\section{Discussion}

\subsection{Provocative question: Where is Meissner effect?}

All known superconductors expel weak external magnetic field. This property is known as the Meissner effect. So why the very strong magnetic field is able to penetrate the QCD superconductor without being suppressed by the superconducting state of the vacuum? The answer to this question is rather simple: the absence of the true Meissner effect is a natural consequence of the anisotropy of the vacuum superconductivity~\eq{eq:London:local}. In fact, the conventional Meissner effect is caused by large superconducting currents which are induced in the superconductor by the external magnetic magnetic field. These currents circulate in the perpendicular (with respect to the magnetic field axis) plane, generating a back-reacting magnetic field which screens the external magnetic field. In the absence of the transverse superconductivity the external magnetic field cannot be screened by the longitudinally superconducting state of the vacuum.

\subsection{A similar effect in condensed matter: Reentrant superconductivity}

Somewhat similar effect, which is known as the magnetic-field-induced, or ``reentrant'', superconductivity, is proposed to be realized in type-II superconductors in a quantum regime~\cite{ref:Tesanovic}. In most superconductors an external magnetic field suppress superconductivity via diamagnetic and Pauli pair breaking effects, so that in a strong magnetic field the superconductivity is lost. However, in a very strong magnetic field the Abrikosov flux lattice of a type-II superconductor may enter a quantum limit of the low Landau level dominance, characterized by the absence of the Meissner effect, a spin-triplet pairing, and a superconducting flow along the magnetic field axis (our proposal~\cite{ref:I,ref:II} of the vacuum superconductivity has exactly the same qualitative features).

Experimentally, the reentrant superconductivity is observed in certain materials. A recent example of such material is a uranium superconductor URhGe, which may exhibit (presumably, spin-triplet) superconductivity at magnetic fields up to 2 Tesla, and comes back to a superconducting state in the region between 8 and 13 Tesla~\cite{ref:Uranium}.

\subsection{Nonperturbative studies of QCD: holography and lattice gauge theory}

Signatures of the $\rho$-meson condensation in QCD were found in nonperturbative holographical approaches based on gauge/gravity duality~\cite{ref:holography}. The spontaneous generation of quark condensates with quantum numbers of electrically charged $\rho$ mesons was also found in numerical simulations of SU(2) lattice gauge theory (quenched QCD)~\cite{ref:lattice}. The critical magnetic field was found to be $B_c =(1.56 \pm 0.13)\cdot 10^{16}\, {\mbox{Tesla}}$ which is quite close to the theoretical expectation in QCD~\eq{eq:eBc}.

\subsection{Nature: Early Universe and heavy-ion collisions}
 
The hadron-scale-strong magnetic fields may emerge in the heavy-ion collisions at the Relativistic Heavy-Ion Collider at Brookhaven National Laboratory and at the Large Hadron Collider at CERN~\cite{ref:Skokov}. Such fields may, presumably, have also arisen in the course of evolution of early Universe~\cite{ref:Universe}, and they may have imprints in the large-scale structure of the magnetic fields in the present-day Universe.


\clearpage

\begin{thebibliography}{99}

\bibitem{ref:I}
  M.~N.~Chernodub,
  { Phys.\ Rev.\  D} {\bf 82}, 085011 (2010)  [arXiv:1008.1055 [hep-ph]].
  
\bibitem{ref:II}  
  M.~N.~Chernodub,
  Phys.\ Rev.\ Lett.\  {\bf 106}, 142003 (2011) [arXiv:1101.0117 [hep-ph]].
 
 \bibitem{ref:Schwinger}
  J.~S.~Schwinger,
  Phys.\ Rev.\  {\bf 82}, 664 (1951).

\bibitem{ref:NO}
  N.~K.~Nielsen, P.~Olesen,  { Nucl.\ Phys.\  B} {\bf 144}, 376 (1978).

\bibitem{ref:AO}
  J.~Ambjorn and P.~Olesen,
  { Nucl.\ Phys.\  B} {\bf 315}, 606 (1989);
  { Int.J. Mod. Phys.}  A {\bf 5}, 4525 (1990);
  { Phys.\ Lett.}  B {\bf 218}, 67 (1989).

\bibitem{ref:Sakurai}
  J.~J.~Sakurai,
  Annals Phys.\  {\bf 11}, 1  (1960).
  
\bibitem{ref:g2}  
  B.~L.~Ioffe and A.~V.~Smilga,
  Nucl.\ Phys.\  B {\bf 232}, 109 (1984);
  A.~Samsonov,
  JHEP {\bf 0312}, 061 (2003) [hep-ph/0308065];
  V.~V.~Braguta and A.~I.~Onishchenko,
  Phys.\ Rev.\  D {\bf 70}, 033001 (2004) [hep-ph/0403258];
  T.~M.~Aliev and M.~Savci,
Phys.\ Rev.\  D {\bf 70}, 094007 (2004) [hep-ph/0405235];
 J.~N.~Hedditch et al,
Phys.\ Rev.\  D {\bf 75}, 094504 (2007) [hep-lat/0703014];
    F.~X.~Lee, S.~Moerschbacher and W.~Wilcox,
Phys.\ Rev.\  D 
  {\bf 78}, 094502 (2008) [arXiv:0807.4150 [hep-lat]].
  
\bibitem{ref:NJL}
  Y.~Nambu and G.~Jona-Lasinio,
  Phys.\ Rev.\  {\bf 122}, 345 (1961).

\bibitem{ref:QED:rho}
  D.~Djukanovic, M.~R.~Schindler, J.~Gegelia and S.~Scherer,
  Phys.\ Rev.\ Lett.\  {\bf 95}, 012001 (2005).
 
\bibitem{ref:isospin}
  D.~N.~Voskresensky,
  Phys.\ Lett.\  B {\bf 392} (1997) 262;
  O.~Aharony, K.~Peeters, J.~Sonnenschein, M.~Zamaklar,
  JHEP {\bf 0802}, 071 (2008)  [arXiv:0709.3948 [hep-th]].

\bibitem{Ammon:2008fc}
  M.~Ammon, J.~Erdmenger, M.~Kaminski, P.~Kerner,
  {Phys.\ Lett.\  B} {\bf 680}, 516 (2009) [arXiv:0810.2316 [hep-th]];
  {JHEP} {\bf 0910}, 067 (2009) [arXiv:0903.1864 [hep-th]].

\bibitem{Abrikosov:1956sx}
  A.~A.~Abrikosov,
  { Sov.\ Phys.\ JETP} {\bf 5}, 1174 (1957)  [Zh.\ Eksp.\ Teor.\ Fiz.\  {\bf 32}, 1442 (1957)].
 
\bibitem{Ebert:1985kz}
  D.~Ebert, H.~Reinhardt,
  Nucl.\ Phys.\  B {\bf 271}, 188 (1986).
  
\bibitem{ref:transition}
  K.~Fukushima, M.~Ruggieri, R.~Gatto,
  Phys.\ Rev.\  D {\bf 81}, 114031 (2010);
  A.~J.~Mizher, M.~N.~Chernodub, E.~S.~Fraga,
Phys.\ Rev.\  D 
{\bf 82}, 105016 (2010);
  M.~D'Elia, S.~Mukherjee, F.~Sanfilippo,
Phys.\ Rev.\  D 
{\bf 82}, 051501 (2010);
  R.~Gatto, M.~Ruggieri,
 Phys.\ Rev.\ D  
{\bf 82}, 054027 (2010);
 Phys.\ Rev.\  D 
{\bf 83}, 034016 (2011);
M.~Ruggieri,
  arXiv:1102.1832, these Proceedings.

\bibitem{ref:catalysis}
  K.~G.~Klimenko,
  Z.\ Phys.\  C {\bf 54}, 323 (1992);
  V.~P.~Gusynin, V.~A.~Miransky and I.~A.~Shovkovy,
  Phys.\ Rev.\ Lett.\  {\bf 73}, 3499 (1994)
  [hep-ph/9405262].
    Nucl.\ Phys.\  B {\bf 462}, 249 (1996) [hep-ph/9509320];
  V.~A.~Miransky and I.~A.~Shovkovy,
  Phys.\ Rev.\  D {\bf 66}, 045006 (2002)
  [hep-ph/0205348].
  
\bibitem{ref:conductivity}  
  P.~V.~Buividovich et al,
  Phys.\ Rev.\ Lett.\  {\bf 105}, 132001 (2010) [arXiv:1003.2180 [hep-lat]];
  P.~V.~Buividovich and M.~I.~Polikarpov,
  arXiv:1011.3001 [hep-lat].
  
\bibitem{ref:lattice}
  V.~V.~Braguta, P.~V.~Buividovich, M.~N.~Chernodub and M.~I.~Polikarpov,
  arXiv:1104.3767 [hep-lat].

\bibitem{ref:Tesanovic}
M.~Rasolt, Phys. Rev. Lett. {\bf 58}, 1482 (1987);
Z.~Te\v{s}anovi\'c, M.~Rasolt, L. Xing {\it ibid.}, {\bf 63} 2425 (1989);
M.~Rasolt, Z.~Te\v{s}anovi\'c, Rev. Mod. Phys. {\bf 64}, 709 (1992).

\bibitem{ref:Uranium}
F. L\'evy, I. Sheikin, B. Grenier, A. D. Huxley, Science {\bf 309}, 1343 (2005); 
D. Aoki, T. D. Matsuda, V. Taufour, E. Hassinger, G. Knebel, J. Flouquet, J. Phys. Soc. Jpn. 78 (2009) 113709; 
arXiv:1012.1987.

\bibitem{ref:holography}
N.~Callebaut, D.~Dudal and H.~Verschelde, PoS {\bf FacesQCD} 046 (2011), arXiv:1102.3103 [hep-ph], these Proceedings;
J.~Erdmenger, P.~Kerner and M.~Strydo, {\it ibid.}  004 (2011), these Proceedings.

\bibitem{ref:Skokov}
  V.~Skokov, A.~Y.~Illarionov and V.~Toneev,
  Int.\ J.\ Mod.\ Phys.\  A {\bf 24}, 5925 (2009);
  V.~Voronyuk et al,
  arXiv:1103.4239 [nucl-th].
  
\bibitem{ref:Universe}
  D.~Grasso and H.~R.~Rubinstein,
  Phys.\ Rept.\  {\bf 348}, 163 (2001) [arXiv:astro-ph/0009061].
 


\end{thebibliography}
\end{document}